\documentclass[prl,twocolumn,showpacs,amssymb,superscriptaddress]{revtex4-1}
\usepackage{amsmath, verbatim, graphicx,color,amsthm}
\usepackage{epsfig}
\usepackage{bbm}
\usepackage[latin1]{inputenc} 

\usepackage{amsfonts}
\usepackage{amssymb}
\usepackage{graphicx}
\newcommand{\bra}[1]{\langle #1|}
\newcommand{\ket}[1]{|#1\rangle}
\newcommand{\ketbra}[1]{| #1\rangle \langle #1|}

\newcommand{\be}{\begin{equation}}
\newcommand{\ee}{\end{equation}}
\newcommand{\eea}{\end{eqnarray}}
\newcommand{\bea}{\begin{eqnarray}}

\newcommand{\XX}{\ensuremath{\mathcal{X}}}

\newcommand{\HH}{\ensuremath{\mathcal{H}}}

\newcommand{\kommentar}[1]{}

\renewcommand{\vr}{\ensuremath{\varrho}}
\newcommand{\forget}[1]{}
\newcommand{\tr}{\mbox{Tr}}

\begin{document}

\title{Estimating entanglement monotones with a generalization of the Wootters formula}
\author{Zhi-Hua Chen}
\affiliation{Department of Science, Zhijiang College, Zhejiang
University of Technology, Hangzhou, 310024,  China}
\author{Zhi-Hao Ma}
\email{ma9452316@gmail.com}
\affiliation{Department of Mathematics, Shanghai Jiaotong University,
Shanghai, 200240, China}
\author{Otfried G{\"u}hne}
\email{otfried.guehne@uni-siegen.de}
\affiliation{Naturwissenschaftlich-Technische Fakult\"at, Universit{\"a}t Siegen,
Walter-Flex-Stra{\ss}e 3, 57068 Siegen, Germany}
\author{Simone Severini}
\affiliation{Department of Computer Science, and Department of Physics \& Astronomy, University College
London, Gower St., WC1E 6BT London, United Kingdom}

\begin{abstract}
Entanglement monotones, such as the concurrence, are useful tools to characterize
quantum correlations in various physical systems. The computation of the concurrence
involves, however, difficult optimizations and only for the simplest case of two
qubits a closed formula was found by Wootters [Phys. Rev. Lett. \textbf{80}, 2245 (1998)].
We show how this approach can be generalized, resulting in lower bounds on the concurrence
for higher dimensional systems as well as for multipartite systems. We demonstrate that
for certain families of states our results constitute the strongest bipartite entanglement
criterion so far; moreover, they allow to recognize novel families of multiparticle bound
entangled states.
\end{abstract}

\date{\today}

\maketitle

{\it Introduction ---}
Entanglement proved itself to be a fundamental concept in physics, with 
applications spanning virtually all areas of quantum science: these 
include antipodal topics such as the black hole information paradox 
and industrial realizations of quantum cryptographic devices. 
By definition, entanglement between two or more particles is given 
by those quantum correlations, which cannot be created by local 
operations or classical communication (LOCC). For the case of more 
than two particles, also different classes of entanglement can be 
distinguished. For the quantification of entanglement and also 
for the discrimination between entanglement classes one can use 
so-called \emph{entanglement measures} or \emph{entanglement monotones} 
-- parameters that indeed are non-increasing under LOCC. The {concurrence}
and the {entanglement of formation} are important parameters of this kind 
\cite{reviews}.

A central problem for the quantification of entanglement is the fact that
nearly all entanglement monotones are extremely difficult to compute. Indeed,
most definitions of entanglement monotones contain nontrivial optimizations,
such as the optimization over all possible LOCC protocols or the minimization
over all possible decompositions of a given density matrix. This difficulty
is an important issue for the application of monotones to real world problems
or experiments.

A milestone in the theory of entanglement measures was the derivation of a
closed formula for the concurrence of two qubits by Wootters in 1998 \cite{Wootters98}.
In this work, it was shown how the minimization over all state decompositions can be
done for such a special case. Consequently, the Wootters' formula has lead to many
applications of the concurrence, \emph{e.g.} for characterizing phase transitions in
spin models \cite{osterlohspin}. In the following years, the formula has been shown
to work also for a special type of multipartite measures by Uhlmann \cite{uhlmann}. 
Furthermore, the concurrence can also be computed for some special states with 
high symmetries \cite{concurrenceresults}.

In the present Letter, we generalize the idea of Wootters to compute lower bounds on
the concurrence. Our methods work for higher dimensional bipartite systems as
well as for multipartite systems. Compared with the large amount of research
about lower bounds on entanglement measures \cite{lowerbounds1, lowerbounds, feibounds} our
approach has substantial advantages: for the bipartite case we discuss a family
of bound entangled states and show that our result gives the strongest
separability criterion so far; for the multipartite examples, our
estimates give the precise value of the multipartite concurrence, and allow
to identify a novel and simple family of bound entangled states. Finally, our
approach can also be used to estimate other quantities besides the concurrence,
which might be useful to deal with entanglement monotones based on antilinear
operators and combs \cite{osterlohsiewert}. It should be noted that lower bounds
on the concurrence based on Wootters' idea have appeared in the literature before
\cite{feibounds}; as we will see, however, the existing approaches are fundamentally 
limited.


{\it Setting the stage ---}
To start, let us recall the main definitions. For a general $m \times n$-dimensional
bipartite pure quantum state $\vr_{AB}=|\psi\rangle\langle\psi|$ on $\mathcal{H}_{A}\otimes\mathcal{H}_{B}$,
the \emph{concurrence} \cite{concurrencearticles, reviews} can be defined as
\begin{equation}
C(|\psi\rangle) = \sqrt{2\left(1-\tr\vr_{A}^{2}\right)  }, \label{con}%
\end{equation}
where $\varrho_{A}=\tr_{B}(|\psi\rangle\langle\psi|)$ is the reduced density
matrix of the first particle \cite{concremark}. A pure state is separable if and only if its concurrence is zero. The above
definitions are extended to mixed states via the so-called \emph{convex roof construction},
\begin{align}
C(\varrho)  &  =\min_{\{p_{i},|\psi_{i}\rangle\}}\sum_{i}p_{i}C(|\psi_{i}%
\rangle),
\end{align}
where the minimization is meant as an optimization over all possible ensemble
realizations $\varrho=\sum_{i}p_{i}|\psi_{i}\rangle\langle\psi_{i}|$, where
$p_{i}\geq0$ and $\sum_{i}p_{i}=1$. The decomposition attaining the minimum
is said to be the \emph{optimal decomposition}. Clearly, this is a
difficult optimization problem, and different estimates have been 
obtained \cite{lowerbounds1, lowerbounds, feibounds}.

{\it The bipartite bound ---}
For our approach, we first need to reformulate the definition of the
concurrence. The pure state $\ket{\psi}$ can be expressed in a product
basis as $\ket{\psi}=\sum_{i=1}^{m}\sum_{j=1}^{n}\psi_{ij}|ij\rangle.$
Furthermore, we can define on $\HH_A$ the generators of the group $SO(m)$
as $L_\alpha = \ket{i}\bra{j}-\ket{j}\bra{i}$. There are $m(m-1)/2$
generators of this type, similarly, there are $n(n-1)/2$ generators $S_\beta$
of $SO(n)$ on $\HH_B.$ Then, a direct calculation for the  $\psi_{ij}$
shows that one can express the concurrence as (see also Ref.~\cite{jpapaper})
\begin{equation}
C^{2}(|\psi\rangle)=
2\left(1-\tr\varrho_{A}^{2}\right)
=\sum_{\alpha,\beta}|\bra{\psi} L_\alpha \otimes S_\beta \ket{\psi^*}|^{2},
\label{aa}%
\end{equation}
where $\ket{\psi^*}$ denotes the complex conjugation. In the following, it is convenient
to use a single index for the matrices $L_{\alpha}\otimes S_{\beta}$ and we define
$J_{t}=L_{\alpha}\otimes S_{\beta}$, where the index $t$ runs from $1$ to $N=[mn(m-1)(n-1)]/4$.

In order to formulate our bound, we first fix an integer $k$. We then choose a subset
of indices $\vec{t}=\{t_{1},...,t_{k}\} \subset \{1,...,N\},$ where we use the ordering
$t_{i}< t_{i+1}$. Moreover, we can choose $k$ complex numbers $\vec{u}=\{u_{s}\}$ for
which the absolute values are bounded via $|u_{s}| \leq 1.$ Then, we consider the quantity
\be
\Delta_k(\varrho, \vec{t},\vec{u})=\max\big\{  0,\lambda_{mn}^{(1)}%
-\sum_{i>1}\lambda_{mn}^{(i)}\big\}  ;
\ee
here the numbers $\lambda_{mn}^{(j)}$ are the square roots of the eigenvalues
of
\be
\XX= \varrho\big(  \sum_{s=1}^{k}u_{s}J_{t_{s}}\big)  \varrho^{\ast}
\big(\sum_{s=1}^{k}u_{s}^{\ast}J_{t_{s}}\big)
\label{eq-bb}
\ee
in non-increasing order. Alternatively, one can say that the
$\lambda_{mn}^{(j)}$ are the eigenvalues of the hermitean matrix
\be
\mathcal{Y}=\sqrt{\sqrt{\varrho} (\sum_{s}u_{s}J_{t_{s}})
\varrho^{\ast}(\sum_{s}u_{s}^{\ast}J_{t_{s}})\sqrt{\varrho}}.
\ee
For our given $k$, we consider the set of {\it all}
possible $\vec{t}$ and choose for any of them a different vector
$\vec{u}$ and compute the corresponding $\Delta_k(\varrho, \vec{t},\vec{u}).$
This leads to ${N}\choose{k}$ numbers and for these we can state our first
main result:

{\bf Observation 1.}
{\it
Let $\varrho$ be a density matrix acting on an $m\times n$-dimensional bipartite
quantum system and consider for fixed $k$ all the possible $\vec{t}$ and a possible
choice of $\vec{u}$ as discussed above. Then, a lower bound on the concurrence
is given by
\begin{equation}
C(\varrho)^{2}
\geq
\frac{N}{k^{2}\binom{N}{k}}\sum_{\vec{t}} [\Delta_k(\varrho, \vec{t},\vec{u})]^2.
\label{conc}
\end{equation}
Especially, if $\varrho$ is separable then $\Delta_k(\varrho, \vec{t},\vec{u})=0$ for any
choice of $k, \vec{t}$ and $\vec{u}.$
}

Before proving this theorem, let us discuss some of its implications.
Eq.~(\ref{conc}) is a lower bound for the concurrence for any given
choice of the $\vec{u}$. In order to obtain a good bound, the set
of the $\vec{u}$ has to be optimized for the given state $\vr$. Often
this has to be done numerically, but we will also present examples,
where a good choice of the $J_{t_{s}}$ is given analytically.

Second, for the case of two qubits there is only one possible
generator, namely $L_\alpha = S_\beta = \ket{0}\bra{1}-\ket{1}\bra{0}=i\sigma_y$.
This implies that the only possibility in Observation 1 is $k=N=1,$ and then
Eq.~(\ref{conc}) reduces to the well known formula for the concurrence of
mixed states. Of course, obtaining a closed formula for the concurrence is a
significantly more advanced result as one has to prove in addition that
equality holds. In Refs.~\cite{Wootters98, uhlmann} this has been achieved by writing down
an explicit decomposition. This is, however, beyond the scope of the present
Letter, we focus on the problem of deriving lower bounds.

Finally, one should add that other researchers have obtained lower bounds
on the concurrence by using the formulation of Eq.~(\ref{aa}) and ideas similar
to the original construction \cite{feibounds}. In these works, the terms
$|\bra{\psi} L_\alpha \otimes S_\beta \ket{\psi^*}|^{2}$ are estimated
separately. A single observable $L_\alpha \otimes S_\beta$, however, acts on
a $2 \times 2$ subspace only, and for these subspaces the criterion of the
positivity of the partial transpose (PPT) is a necessary and sufficient criterion
for entanglement \cite{reviews}. This implies that the approaches in Refs.~\cite{feibounds} can
never detect weak forms of entanglement, such as bound entanglement which is not detected by
the PPT criterion \cite{feiremark}. On the other side, Observation 1, represents a strong
criterion for bound entanglement, as we will see below.

{\it Proof of Observation 1.}
First we prove that for a fixed $k$, and fixed vector $\vec{t}$ we have that
\begin{equation}
\min_{\{p_{i},|\psi_{i}\rangle\}}\big\{\sum_{i}%
p_{i}|\langle\psi_{i}|\sum_{s=1}^{k}u_{s}J_{t_{s}}|\psi_{i}^*%
\rangle|\big\} \geq \Delta_k(\varrho, \vec{t},\vec{u}),
\end{equation}
where the minimum is taken over all decompositions
$\varrho=\sum_{i}p_{i}|\psi_{i}\rangle\langle\psi_{i}|$.
Let $\lambda_{i}$ and $|\chi_{i}\rangle$ be the eigenvalues and the eigenvectors of
$\varrho$. It is known that any decomposition of $\varrho$ is connected to the
eigenvalue decomposition via a unitary matrix $U_{ij}$, namely one has
$\sqrt{p_i}|\psi_{i}\rangle=\sum_{j=1}^{mn}U_{ij}^{\ast}(\sqrt{\lambda_{j}}|\chi_{j}\rangle)$
\cite{nielsenbook}.
Therefore, we have
$\sqrt{p_i p_j}\langle\psi_{i}|\sum_{s=1}^{k}u_{s}J_{t_{s}}|\psi_{j}^{\ast}\rangle
=(UYU^{T})_{ij}$,
where the matrix $Y$ is defined by
$Y_{\alpha \beta}= \sqrt{\lambda_\alpha \lambda_\beta}
\langle\chi_{\alpha}|\sum_{s=1}^{k}u_{s}J_{t_{s}}|\chi_{\beta}^{\ast}\rangle$.
Since the $J_k$ are symmetric, the matrix $Y=Y^T$ is complex and symmetric and
we can use Takagi's factorization \cite{takagi} to write $Y=V D V^T$ with
a real diagonal matrix $D.$  The entries of $D$ are nonnegative and given by
the square roots of the eigenvalues of $YY^{\dagger}$. Then, following directly
the argumentation of Ref.~\cite{Wootters98} we have:
\begin{align}
& \min_{\{p_{i},|\psi_{i}\rangle\}}
\big\{  \sum_{i}p_{i}|\langle\psi_{i}|\sum_{s=1}^{k}%
u_{s}J_{t_{s}}|\psi_{i}^*\rangle|\big\}
\nonumber
\\
&
=\min_{W=UV}
\big\{\sum_{i}|[WDW^{T}]_{ii}|\big\}
\geq \lambda_{mn}^{(1)}-\sum_{i>1}\lambda_{mn}^{(i)},
\nonumber
\\
& = \Delta_k(\varrho, \vec{t},\vec{u})
\end{align}
where $\lambda_{mn}^{(j)}$ are the entries of $D$ in decreasing order. These
quantities are, however, nothing but the eigenvalues of $\mathcal{X}$ in
Eq.~(\ref{eq-bb}). Therefore, if a state $\varrho$ is separable then a
decomposition into pure states without concurrence exists. Due to Eq.~(\ref{aa})
all the mean values of $J_k$ vanish, which implies already that
$\Delta_k(\varrho, \vec{t},\vec{u})=0.$

It remains to show that $\Delta_k(\varrho, \vec{t},\vec{u})$ can give a lower
bound on the concurrence also for entangled states. Suppose that
$\varrho=\sum_{i}p_{i}|\psi_{i}\rangle\langle\psi_{i}|$ is an optimal
decomposition of $\varrho$. Then
$
C(\varrho)   = \sum_{i}p_{i}C(|\psi_{i}\rangle)
  =\sum_{i}p_{i}\sqrt{\sum_{t=1}^N|\bra{\psi_i} J_t \ket{\psi_i^*}|^{2}}.
$
From the argumentation above, we know that for fixed $k$ and $\vec{t}$ and
fixed $t_{1},...,t_{k}$ the estimates
$
\Delta_k(\varrho, \vec{t},\vec{u}) \leq
\sum_{i}p_{i}\sum_{s=1}^{k}|\langle\psi_{i}|u_s J_{t_{s}}|\psi_{i}^*\rangle|
\leq
\sum_{i}p_{i}\sum_{s=1}^{k}|\langle\psi_{i}|J_{t_{s}}|\psi_{i}^*\rangle|
$
hold.

Finally, using the rule $(\sum_{j=1}^k x_j)^2 \leq k \sum_j x_j^2 $ and the
Cauchy-Schwartz inequality we can directly estimate the right-hand side
of Eq.~(\ref{conc}) as:
\begin{align}
&\sum_{\vec{t}} [\Delta_k(\varrho, \vec{t},\vec{u})]^2  \leq
\frac{k^{2}}{N}{N\choose k} C(\varrho)^{2}.
\end{align}
The details of this calculation are given in the Appendix A1 \cite{suppmat}. This concludes the
proof of Observation 1.
\qed

Before proceeding to the examples, let us discuss the properties of
the concurrence that were used in the proof. The starting point was
Eq.~(\ref{aa}) and the only further requirement needed was that the
fact that the $J_t=J_t^T$ were symmetric \cite{symmremark}. Moreover,
if ${A}_t=-{A}_t^T$ were antisymmetric, then one has for any
state $|\bra{\psi}A_t\ket{\psi^*}|^2 = 0$, so restricting to
symmetric $J_t$ can be done without loosing generality. In summary,
the convex roof of any quantity $E(\ket{\psi})$, which can be written
as
\be
E^2(\ket{\psi})= \sum_t \pm m_t |\bra{\psi}M_t \ket{\psi^*}|^2,
\label{genform}
\ee
can be estimated with our methods: one can split each $M_t$ in
a symmetric and an antisymmetric part and estimate the contributions
from the symmetric part. The fact that some of the coefficients $m_t$
can be negative does not matter: using the relation
$\sum_{t}|\bra{\psi} G_t \ket{\psi^*}|^2 = 1$
(where the $G_t$ form an orthonormal basis of the operator space)
one can rewrite $E^2(\ket{\psi})$ as a sum with only positive
coefficients minus a constant term.

{\it Bound entangled states as an example ---}
In order to show that Observation 1 results in a stronger separability 
criterion than best methods that are currently known, we consider the 
family of $3\times3$ bound entangled states introduced by P. Horodecki 
\cite{Horodeckibound}.  This family of states $\varrho_{a}^{PH}$ is 
not detected by PPT criterion, but is nevertheless entangled for any 
$0<a<1$. The detailed form of these states is given in Appendix A2 \cite{suppmat}. 
We consider a mixture of these states with white noise, 
$\varrho_{a}(p)=p\varrho^{PH}_a+(1-p){\openone}/9$ and ask for the minimal 
$p$, so that the entanglement in $\varrho_{a}(p)$ is still detected. 
First, we use Observation 1 with the purpose of detecting entanglement 
and find the optimal $J_t$ via numerical optimization. We finally compare 
our values with the values obtained via different known criteria: 
the Zhang-Zhang-Zhang-Guo (ZZZG) criterion \cite{CJ10}, the Ma and Bao 
(MB) criterion \cite{Ma10}, and the method based on symmetric extensions 
and semidefinite programming (SDP) \cite{sdp, critremark}. We also used the 
algorithm proposed in Ref.~\cite{sepalgo} to prove separability of 
quantum states. This allows to compute values of $p$, for which 
$\varrho_{a}(p)$ is provably separable.

\begin{figure}[t]
\begin{center}
\includegraphics[width=\columnwidth]{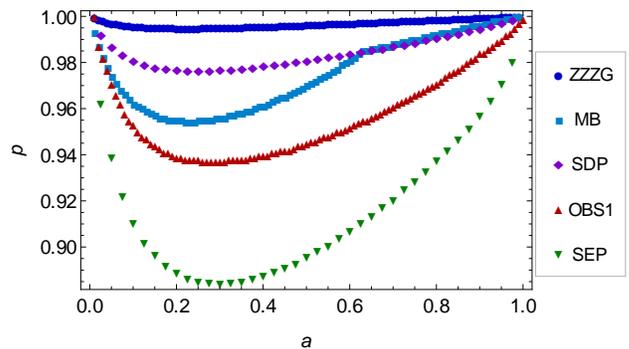}
\caption{(Color online) Detecting entanglement in the 
Horodecki $3\times3$ bound entangled state mixed with 
white noise. The criterion of Observation 1 (points 
denoted by OBS1) is stronger than previously known 
criteria. For values of $p$ smaller than the  values 
given by SEP the states $\varrho_{a}(p)$ are proven 
to be separable. See text for further details.
}
\label{fig2}%
\end{center}
\end{figure}

The results are given in Fig. 1. One clearly sees that Observation 1
provides the best criterion, but the comparison with the separability
algorithm also suggests that Observation 1 does not detect all states.

{\it Estimating the multipartite concurrence ---}
For simplicity, we only discuss the three particle case, but our 
results can be directly generalized to arbitrary $N$-partite states.
Let us consider a pure state $\ket{\psi}$ in a $d\times d \times d$-system.
Its concurrence is given by
\begin{equation}
C^{\tau}(|\psi\rangle) = \sqrt{[3-(\tr\varrho_1^2+\tr\varrho_2^2+\tr\varrho_3^2)]},
\end{equation}
where $\varrho_1=\tr_{23}(\varrho)$, \emph{etc.} are the reduced one-particle states.
For this definition, it directly follows that for pure states
$
C^{\tau}(|\psi\rangle)^2=\frac{1}{2}
([C^{(1|23)}(|\psi\rangle)]^2+
[C^{(2|13)}(|\psi\rangle)]^2+
[C^{(3|12)}(|\psi\rangle)]^2
),
$
where $C^{(1|23)}(|\psi\rangle)$, \emph{etc.} are the corresponding bipartite
concurrences. This definition is extended to mixed states via the convex
roof construction. Clearly, $C^{\tau}(\varrho)=0$ if and only if $\varrho$
is a fully separable state.

A first possibility to estimate the multipartite concurrence is to start with
an estimate of the bipartite concurrence for each bipartition (as in Observation 1),
and then estimate the total concurrence $C^{\tau}$ from it. This is indeed a
viable way, in Appendix A3 \cite{suppmat} we present and discuss a corresponding theorem.
The disadvantage of this approach is that there are states which are separable for
any bipartition, but not fully separable \cite{fullsep}. For them, this method will not
succeed, since all the bipartite concurrences vanish.

To overcome this limitation, one should note that $C^{\tau}(|\psi\rangle)^2$ is of the same
structure as Eq.~(\ref{genform}): we define the operators $J_t$ as before, but separately
for any bipartition and write
$J^{1|23}_t=L^1_\alpha \otimes S_\beta^{23}$ and similarly for the other bipartitions.
Then we have the expression
$C^{\tau}(|\psi\rangle)^2:= \frac{1}{2}\sum_{t}
[|\langle\psi|J_t^{1|23}|\psi^*\rangle|^2
+|\langle\psi|J_t^{2|13}|\psi^*\rangle|^2
+|\langle\psi|J_t^{3|12}|\psi^*\rangle|^2]$.
So we have to consider
\be
\Delta^{tot}_k(\varrho, \vec{t}, \vec{x})= \max{(0,\lambda_{mn}^{(1)}-\sum_{i>1}\lambda_{mn}^{(i)})},
\ee
where the $\lambda_{mn}^{(j)}$ are the square roots of eigenvalues of
\begin{align}
\mathcal{X}^{tot}=&\varrho
\sum_{s=1}^k (u_{s}J_{t_s}^{1|23}+v_{s}J_{t_s}^{2|13}+w_{s}J_{t_s}^{3|12})
\varrho^*
\nonumber
\\
&
\times
\sum_{s=1}^k (u^{*}_{s}J_{t_s}^{1|23}+v^{*}_{s}J_{t_s}^{2|13}+w^{*}_{s}J_{t_s}^{3|12})
\end{align}
in decreasing order. Here, $\vec{x}=(\vec{u},\vec{v},\vec{w})$ denotes a triple of complex
vectors which are normalized as in Observation 1 and $\vec{t}=\{t_1,...,t_k\}$. For this
quantity we can state the following:

{\bf Observation 2.}
{\it
For any arbitrary mixed state on $\mathcal{H}\otimes \mathcal{H}\otimes \mathcal{H}$
and for every fixed $k$ and for arbitrary $\vec{x}$ we have:
\be
\frac{N}{6k^{2} {N\choose k}} \sum_{\vec{t}}[(\Delta^{tot}_k(\varrho, \vec{t},\vec{x}))^2] \leq C^\tau(\varrho)^2.
\label{th3eq}
\ee
}
A proof is given in the Appendix A4 \cite{suppmat}.

{\it Multipartite Examples ---}
 We will
consider two simple examples for three qubits, but these already demonstrate
two interesting points: first, they give an idea how the observables $J_t$ and the
coefficients $\vec{x}$ can be chosen; second, it turns out that the entanglement
criterion in Observation 2 is strong and allows to identify a novel family
of bound entangled states.

As the first example, we consider the three-qubit Greenberger-Horne-Zeilinger (GHZ)
state
$\ket{GHZ}=(\ket{000}+\ket{111})/{\sqrt{2}}$
and mix it with white noise, $\varrho^G(p)=p \ketbra{GHZ}+(1-p)\openone/8.$
Then we take the single-qubit operator $S^{(a)}=|0\rangle\langle1|-|1\rangle\langle0|$
and the two-qubit operator $L^{(bc)}=|00\rangle\langle11|-|11\rangle\langle00|$ and
from them we form the operators $J^{i|jk}=S^{(i)}\otimes L^{(jk)}$ for all three
bipartitions. Applying Observation 2 for the choice $k=1$ and $u_1 = v_1 = w_1 = 1$,
one finds already from a single term in the sum of Eq.~(\ref{th3eq}) that the
three-qubit concurrence is bounded by
\be
(C^\tau[\varrho^G(p)])^2 \geq \frac{1}{6}\big(\frac{3}{4}[5p-1]\big)^2.
\label{multiex1}
\ee
For $p=1$, this reproduces exactly concurrence of the pure GHZ state. Moreover,
this bound shows that the state $\varrho^G(p)$ is entangled for $p > 1/5.$ This
means that Observation 2 provides a necessary and sufficient criterion for entanglement
for the family of states $\varrho^G(p)$, since it is known that for $p\leq 1/5$
these states are separable \cite{ghzschranke}. In fact, Eq.~(\ref{multiex1}) gives a linear
lower bound on the convex function $C^\tau[\varrho^G(p)]$ and this bound coincides with
the exact value on the points $p=1/5$ and $p=1.$ This means that the bound equals
the exact value on the whole interval $p\in [1/5; 1]$ and for them we have
$C^\tau[\varrho^G(p)] = (\tfrac{3}{4}[5p-1])/\sqrt{6}.$

As the second example, we consider the three-qubit W state
$\ket{W}=(\ket{001}+\ket{010}+\ket{100})/{\sqrt{3}}$
mixed with white noise, $\varrho^W(p)=p \ketbra{W}+(1-p)\openone/8.$
In this  case, we use again the operator $S^{(a)}=|0\rangle\langle1|-|1\rangle\langle0|$,
but for two qubits we use the $L^{(bc)}=|00\rangle\langle10|-|10\rangle\langle00|$
and from them we form the operators $J^{i|jk}=S^{(i)}\otimes L^{(jk)}.$ Applying Observation 2
for $k=1$ and $u_1 = v_1 = w_1 = 1$, we find that $(C^\tau[\varrho^W(p)])^2 \geq (1/96)[p(8+\sqrt{3})-\sqrt{3}]^2$,
especially, the state $\varrho^W(p)$ is entangled for $p > p_s = \sqrt{3}/(8+\sqrt{3}) \approx 0.178.$

This is a remarkable value for several reasons. First, using the separability
algorithm from Ref.~\cite{sepalgo}, one can prove that the states $\varrho^W(p)$ 
are fully separable for $p \leq  0.177,$ giving strong evidence that Observation 2 
provides a necessary and sufficient criterion for the family of states $\varrho^W(p).$

Second, these calculations show that the states $\varrho^W(p)$ exhibit quite peculiar 
entanglement properties: one can directly check that for 
$p \leq 3(8\sqrt{2}-3)/119 \approx 0.2096$ the states have a positive partial 
transpose for any bipartition, and using the separability algorithm \cite{sepalgo} 
one finds that for $p \leq 0.2095 $ the states are indeed separable for any bipartition. 
Hence, for $p \in [p_s; 0.2095]$ the states $\varrho^W(p)$ are separable for any bipartition,
but not fully separable. This implies that they are bound entangled: no entanglement
can be distilled from them, even if two of the three parties join. It was known that
such states exist \cite{fullsep}, however, the existing examples required a sophisticated
construction. It is surprising that the simple family $\varrho^W(p)$
includes bound entangled states and it underlines the power of our approach that
these states can be detected with Observation 2. Finally, the bound entanglement
in the family $\varrho^W(p)$ can easily be generated experimentally (contrary to
other known examples of bound entangled states) since adding noise to a pure state
is easy in practically any experimental implementation. 

{\it Conclusion ---} We have provided a general method to bound entanglement
monotones by extending in a nontrivial way the original construction of Wootters 
\cite{Wootters98}, an approach that works for both bipartite and multipartite 
concurrence. We leave open the problem of determining for which states our 
method gives the exact value of the concurrence. It would also be interesting 
to broaden our approach to the general classification of invariants of quantum 
states \cite{osterlohsiewert}, since this may help to understand the different
entanglement classes for multiparticle systems.

We thank M. Hofmann for discussions.
Z. C. is supported by NSF of China (11201427),
Z. M. is  supported by the NSF of China (10901103), 
O. G. is supported by the EU (Marie Curie CIG 293993/ENFOQI) 
and the BMBF (Chist-Era Project QUASAR), and S. S. is supported 
by the Royal Society.

\newpage

\begin{widetext}

\section{Appendix}

\subsection*{A1: Details of the calculation in the proof of Observation 1}
Let us write $\Gamma_{k}^{(i)}= |\bra{\psi_i} J_{t_k} \ket{\psi_i^*}|$. By making use of the
rule $(\sum_{j=1}^k x_j)^2 \leq k \sum_{j=1}^k x_j^2 $ and the Cauchy-Schwartz inequality
we can estimate the following:
\begin{align}
\sum_{\vec{t}} [\Delta_k(\varrho, \vec{t},\vec{u})]^2 &  \leq
\sum_{\vec{t}} \Big(\sum_{i}p_{i}\sum_{s=1}^{k}\Gamma_{s}^{(i)}\Big)^{2}
\nonumber
\\
&  =\sum_{i}p_{i}^{2}\sum_{\vec{t}}\Big(\sum_{s=1}^{k}\Gamma_{s}^{(i)}\Big)^{2}
+2\sum\limits_{i<j}p_{i}p_{j}
\sum_{\vec{t}}\sum_{h=1}^{k}\Gamma_{h}^{(i)}\sum_{m=1}^{k}\Gamma_{m}^{(j)}
\nonumber
\\
&  \leq \sum_{i} p_{i}^{2}\sum_{\vec{t}} k \sum_{s=1}^{k} (\Gamma_{s}^{(i)})^{2}
+2\sum\limits_{i<j}p_{i}p_{j}
\sqrt{\sum_{\vec{t}}\Big(\sum\limits_{h=1}^{k}\Gamma_{h}^{(i)}\Big)^{2}}
\sqrt{\sum_{\vec{t}}\Big(\sum\limits_{m=1}^{k}\Gamma_{m}^{(j)}\Big)^{2}}
\nonumber
\\
&  \leq k \sum_{i} p_{i}^{2}\sum_{\vec{t}}  \sum_{s=1}^{k} (\Gamma_{s}^{(i)})^{2}
+2k\sum_{i<j}p_{i}p_{j}
\sqrt{\sum_{\vec{t}} \sum_{h=1}^{k} (\Gamma_{h}^{(i)})^2}
\sqrt{\sum_{\vec{t}} \sum_{m=1}^{k} (\Gamma_{m}^{(j)})^2}
\nonumber
\\
&  = \frac{k^{2}}{N}{N\choose k} \sum_{i} p_{i}^{2} \sum_{t=1}^{N}
(\Gamma_{t}^{(i)})^{2}
+\frac{2 k^{2}}{N} {N\choose k} \sum_{i<j}p_{i}p_{j}
\sqrt{\sum_{t=1}^{N}(\Gamma_{t}^{(i)})^{2}}
\sqrt{\sum_{t=1}^{N}(\Gamma_{t}^{(j)})^{2}}
\nonumber
\\
&  =\frac{k^{2}}{N}{N\choose k} \Big(\sum_{i}p_{i}\sqrt{\sum_{t=1}^{N}
(\Gamma_{t}^{i})^{2}}\Big)^{2}
\nonumber
\\
&  =\frac{k^{2}}{N}{N\choose k} C(\varrho)^{2}.
\end{align}
\qed

We would like to add that for the case $k=N$ also a different bound can be proved: 
Consider $\Delta^{\rm tot}(\vr,\vec{u}) = \Delta_N(\vr,\vec{t},\vec{u})$ where $\vec{u}$ 
is now normalized as a vector, that is $\sum_{s=1}^N u_s^* u_s =1.$ Note that since 
$k=N$, $\vec{t}$ is fixed and denotes all matrices $J_t.$ Then, one has that
\be
C(\vr) \geq \Delta^{\rm tot}(\vr,\vec{u}).
\ee
This can be seen as follows. First, Eq. (8) in the main text can be proven just as before.
Then, we have for the optimal decomposition
$
\Delta^{\rm tot}(\vr,\vec{u})
\leq 
\sum_i p_i \sum_{t=1}^{N} |\bra{\psi_i} u_t J_t \ket{\psi_i^*}|
=
\sum_i p_i \sum_{t=1}^{N} |u_t| |\bra{\psi_i} J_t \ket{\psi_i^*}|
\leq 
\sum_i p_i \sqrt{ \sum_{t=1}^{N} |\bra{\psi_i} J_t \ket{\psi_i^*}|^2}
= C(\vr),
$
where also the Cauchy-Schwartz inequality has been used.

\subsection*{A2: The family of bound entangled states}
The family of bound entangled states from Ref.~\cite{Horodeckibound} are explicitly
given by
\be
\varrho^{PH}_{a}=\frac{1}{8a+1}
\begin{pmatrix}
a & 0 & 0 & 0 & a & 0 & 0 & 0 &a
\\
0 & a & 0 & 0 & 0 & 0 & 0 & 0 &0
\\
0 & 0 & a & 0 & 0 & 0 & 0 & 0 &0
\\
0 & 0 & 0 & a & 0 & 0 & 0 & 0 &0
\\
a & 0 & 0 & 0 & a & 0 & 0 & 0 &a
\\
0 & 0 & 0 & 0 & 0 & a & 0 & 0 &0
\\
0 & 0 & 0 & 0 & 0 & 0& \tfrac{1+a}{2}  & 0 & \tfrac{\sqrt{1-a^2}}{2}
\\
0 & 0 & 0 & 0 & 0 & 0 & 0 & a &0
\\
a & 0 & 0 & 0 & 0 &  0 & \tfrac{\sqrt{1-a^2}}{2}  & 0 &\tfrac{1+a}{2}
\end{pmatrix}.
\ee
These states have a non-negative partial transpose and are not distillable,
but they are entangled for any $0<a<1$.

\subsection*{A3: Estimates on the multipartite concurrence from bipartite
bounds}
We define the operators $J_k$ as before, but separately for any bipartition.
That is, we write $J^{1|23}_k=L^1_\alpha \otimes S_\beta^{23}$ and similarly
for the other bipartitions. Then, we obtain the expression
$C^{\tau}(|\psi\rangle)^2:= \frac{1}{2}\sum_{k}
[|\langle\psi|J_k^{1|23}|\psi^*\rangle|^2
+|\langle\psi|J_k^{2|12}|\psi^*\rangle|^2
+|\langle\psi|J_k^{3|12}|\psi^*\rangle|^2]$.
Now, let us consider
\be
\Delta_k^{1|23}(\varrho, \vec{t},\vec{u})=\max\left\{  0,\lambda_{mn}^{(1)}%
-\sum\nolimits_{i>1}\lambda_{mn}^{(i)}\right\},
\ee
where the numbers $\lambda_{mn}^{(j)}$ are the square roots of the eigenvalues
of
$
\varrho\big(\sum_{s=1}^{k}u_{s}J^{1|23}_{t_{s}}\big)  \varrho^{\ast}
\big(\sum_{s=1}^{k}u_{s}^{\ast}J^{1|23}_{t_{s}}\big)
$
in non-increasing order and similarly for the other bipartitions.
We can then state:

{\bf Observation 3.}
{\it
Let $\varrho$ be a density matrix on a tripartite $d\times d\times d$-system and
consider for fixed $k$ all the possible $\vec{t}$ and a possible choice of $\vec{u}$
as in Observation 1. Then, a lower bound on the multipartite concurrence is given by
\begin{align}
C^{\tau}(\rho)^{2}
\geq
\frac{1}{2}
\frac{N}{k^{2}\binom{N}{k}}\sum_{\vec{t}}
\big\{
&
[\Delta_k^{1|23}(\varrho, \vec{t},\vec{u})]^2
+
[\Delta_k^{2|13}(\varrho, \vec{t},\vec{u})]^2
+
[\Delta_k^{3|12}(\varrho, \vec{t},\vec{u})]^2
\big\}.
\label{concmult}%
\end{align}
Here, the coefficients $\vec{u}$ can be chosen separately for any $\vec{t}$ and any bipartition.
Especially, if $\varrho$ is fully separable then $\Delta_k^{a|bc}(\varrho, \vec{t},\vec{u})=0$
for any choice of $k, \vec{t}$ and $\vec{u}.$}

{\it Proof.}
First, one finds in the same way as in the bipartite case:
\begin{equation}
\frac{N}{k^{2}{N\choose k}}
\sum_{\vec{t}}[\Delta^{1|23}(\varrho, \vec{t}, \vec{u})]^2
\leq
[C^{1|23}(\varrho)]^2
\end{equation}
and analogous bounds for the other bipartitions. It remains to bound $C^{\tau}(\varrho)$
from the values of $C^{1|23}(\varrho), C^{2|13}(\varrho)$ and $C^{3|12}(\varrho)$.
For that, let us assume that $\varrho=\sum_i p_i \ketbra{\psi_i}$ is an optimal
decomposition when computing the convex roof of $C^\tau(\varrho).$  Let us denote
$\Theta^{(i)(a|bc)}_k=|\bra{\psi_i}J_k^{a|bc}\ket{\psi_i^*}|^2$. Using the
Cauchy-Schwartz inequality we have that
\begin{align}
[C^{1|23}&(\varrho)]^2  + [C^{2|13}(\varrho)]^2 + [C^{3|12}(\varrho)]^2
\nonumber
\\
&\leq
\Big(
\sum_i p_i \sqrt{\sum_k \Theta^{(i)(1|23)}_k}
\Big)^2
+
\Big(
\sum_i p_i \sqrt{\sum_k \Theta^{(i)(2|13)}_k}
\Big)^2
+
\Big(
\sum_i p_i \sqrt{\sum_k \Theta^{(i)(3|12)}_k}
\Big)^2
\nonumber
\\
&=
\sum_i \Big\{ p_i^2 \sum_k (\Theta^{(i)(1|23)}_k+\Theta^{(i)(2|13)}_k+\Theta^{(i)(3|12)}_k\Big\}
\nonumber
\\
&
+ 2\sum_{i<j} \Big\{ p_i p_j
\sqrt{\sum_k \Theta^{(i)(1|23)}_k}\sqrt{\sum_k \Theta^{(j)(1|23)}_k}+
\sqrt{\sum_k \Theta^{(i)(2|13)}_k}\sqrt{\sum_k \Theta^{(j)(2|13)}_k}
\nonumber
\\
&+
\sqrt{\sum_k \Theta^{(i)(3|12)}_k}\sqrt{\sum_k \Theta^{(j)(3|12)}_k}
\Big\}
\nonumber
\\
&\leq
\sum_i \Big\{ p_i^2 \sum_k (\Theta^{(i)(1|23)}_k+\Theta^{(i)(2|13)}_k+\Theta^{(i)(3|12)}_k\Big\}
\nonumber
\\
&
+2\sum_{i<j} \Big\{p_i p_j
\sqrt{\sum_k \Theta^{(i)(1|23)}_k+\Theta^{(i)(2|13)}_k+\Theta^{(i)(3|12)}_k}
\sqrt{\sum_k \Theta^{(i)(1|23)}_k+\Theta^{(i)(2|13)}_k+\Theta^{(i)(3|12)}_k}
\Big\}
\nonumber
\\
&
= 2[\sum_i p_i C^{\tau}(\ket{\psi_i})]^2 =2C^{\tau}(\varrho)^2,
\end{align}
which proves the claim.
\qed

\subsection*{A4: Proof of Observation 2}
First, as in Observation 1, we can prove that
\begin{align}
\min_{\{p_{i},|\psi_{i}\rangle\}}
\left\{  \sum\nolimits_{i}p_{i}|\langle\psi_{i}|\sum_{s=1}^{k}%
(u_{s}J^{1|23}_{t_{s}}+v_{s}J^{2|13}_{t_{s}}+w_{s}J^{3|12}_{t_{s}})|\psi^{*}_{i}\rangle|\right\}
& \geq \Delta^{tot}_k(\varrho, \vec{t},\vec{x})
\end{align}
By denoting $\Theta^{(i)(a|bc)}_k=|\bra{\psi_i}J_k^{a|bc}\ket{\psi_i^*}|^2$,
\begin{align}
\sum_{\vec{t}} & (\Delta^{tot}_k(\varrho, \vec{t},\vec{x}))^2
\leq
\sum_{\vec{t}}
\Big[
\sum_{i} p_{i} \big(|\langle\psi_{i}|\sum_{s=1}^{k} J^{1|23}_{t_{s}}|\psi_{i}^{*}\rangle|+
|\langle\psi_{i}|\sum\limits_{s=1}^{k}J^{2|13}_{t_{s}}|\psi_{i}^{*}\rangle|+|\langle\psi_{i}|\sum\limits_{s=1}^{k}%
J^{3|12}_{t_{s}}|\psi_{i}^{*}\rangle|\big)\Big]^2
\nonumber
\\
&=\sum_{\vec{t}}
\Big[
\sum_{i}p_{i}\sum\limits_{s=1}^{k}
\big(|\sqrt{\Theta^{(i){(1|23)}}_{t_s}}|+
|\sqrt{\Theta^{(i)(2|13)}_{t_s}}|+
|\sqrt{\Theta^{(i){(3|12)}}_{t_s}|}\big)
\Big]^2
\nonumber
\\
&
\leq \sum_{i} p_{i}^{2}
\sum_{\vec{t}} k
\sum_{s=1}^{k} \big(\Theta_{t_s}^{(i)(1|23)}+\Theta_{t_s}^{(i)(2|13)}+\Theta_{t_s}^{(i)(3|12)}
+2\sqrt{\Theta_{t_s}^{(i)(1|23)}\Theta_{t_s}^{(i)(2|13)}}
+2\sqrt{\Theta_{t_s}^{(i)(1|23)}\Theta_{t_s}^{(i)(3|12)}}
\nonumber
\\
&+2\sqrt{\Theta_{t_s}^{(i)(2|13)}\Theta_{t_s}^{(i)(3|12)}}\big)
+2\sum\limits_{i<j}p_{i}p_{j}
\sum_{\vec{t}}
\Big(\sum\limits_{s=1}^{k}\sqrt{\Theta_{t_s}^{(i)(1|23)}}+\sqrt{\Theta_{t_s}^{(i)(2|13)}}+\sqrt{\Theta_{t_s}^{(i)(3|12)}}\Big)
\nonumber
\\
&
\times \Big(\sum\limits_{s=1}^{k}\sqrt{\Theta_{t_s}^{(j)(1|23)}}+\sqrt{\Theta_{t_s}^{(j)(2|13)}}+\sqrt{\Theta_{t_s}^{(j)(3|12)}}\Big)
\nonumber
\\
&
\leq \sum_{i} p_{i}^{2}
\sum_{\vec{t}} k \sum_{s=1}^{k} (3\Theta_{t_s}^{(i)(1|23)}+3\Theta_{t_s}^{(i)(2|13)}+3\Theta_{t_s}^{(i)(3|12)}
\nonumber
\\
&+2\sum_{i<j}p_{i}p_{j}
3 k
\sqrt{\sum_{\vec{t}}\sum_{s=1}^{k}(\Theta_{t_s}^{(i)(1|23)}+\Theta_{t_s}^{(i)(2|13)}+\Theta_{t_s}^{(i)(3|12)})}
\sqrt{\sum_{\vec{t}}\sum_{s=1}^{k}(\Theta_{t_s}^{(j)(1|23)}+\Theta_{t_s}^{(j)(2|13)}+\Theta_{t_s}^{(j)(3|12)})}
\nonumber
\\
&
=
\frac{3k^2}{N}{N\choose k}
\Big[\sum\limits_{i} p_{i}^{2} \sum\limits_{s=1}^{N} (\Theta_{t_s}^{(i)(1|23)}+\Theta_{t_s}^{(i)(2|13)}+\Theta_{t_s}^{(i)(3|12)})
\nonumber
\\
&+2\sum\limits_{i<j}p_{i}p_{j}\sqrt{\sum\limits_{s=1}^{N}(\Theta_{t_s}^{(i)(1|23)}+\Theta_{t_s}^{(i)(2|13)}+\Theta_{t_s}^{(i)(3|12)})}
\sqrt{\sum\limits_{s=1}^{N}(\Theta_{t_s}^{(j)(1|23)}+\Theta_{t_s}^{(j)(2|13)}+\Theta_{t_s}^{(j)(3|12)})}
\Big]
\nonumber
\\
&
=\frac{6k^2}{N}{N\choose k}[\sum_i p_i C^{\tau}(\ket{\psi_i})]^2
\nonumber
\\
& = \frac{6k^2}{N}{N\choose k}C^{\tau}(\varrho)^2.
\end{align}
This concludes the proof. 
\qed

\end{widetext}


\begin{thebibliography}{99}
\bibitem{concremark}
Note that originally the concurrence was expressed differently via 
a spin-flip operation \cite{concurrencearticles, Wootters98}.
\bibitem{feiremark}
This is also explicitely stated in Ref.~\cite{feibounds} (b).
\bibitem{symmremark}
Of course, for any observable one can always find a basis where it is
symmetric, but in the present case this basis has to coincide with the
basis where the complex conjugation $\ket{\psi^*}$ is taken.
\bibitem{critremark}
We also considered the algebraic bound from Ref.~\cite{lowerbounds1}(a) 
the quasipure approximation from Ref.~\cite{lowerbounds1}(b), the computable 
cross-norm or realignment (CCNR) criterion and the covariance matrix criterion.
The bounds from Ref.~\cite{lowerbounds1} are of a similar strength as the SDP 
criterion, e.g. the algebraic bound from Ref.~\cite{lowerbounds1}(a) results
for $a=0.2$ in $p=0.968$ and the bound from Ref.~\cite{lowerbounds1}(b) in 
$p=0.971.$ The other criteria are  weak in comparison with the SDP.
\end{thebibliography}
\end{document}